\documentclass[10pt,preprint]{aastex}
\slugcomment{Version: \today}

\usepackage{color}
\usepackage{graphicx}

\begin{document}
\title{Continuum Variability of Deeply Embedded Protostars as a Probe of Envelope Structure}
\author{
        Doug Johnstone\altaffilmark{1,2,3},
        Benjamin Hendricks\altaffilmark{3,4},        
        Gregory J. Herczeg\altaffilmark{5},
        \& Simon Bruderer\altaffilmark{6}
}
\altaffiltext{1}{Joint Astronomy Centre, 660 North A'ohoku Place, University Park, Hilo, HI 96720, USA}
\altaffiltext{2}{National Research Council Canada, Herzberg
Institute of Astrophysics, 5071 West Saanich Rd, Victoria, BC, V9E
2E7, Canada; doug.johnstone@nrc-cnrc.gc.ca}
\altaffiltext{3}{Department of Physics \& Astronomy, University of Victoria,
Victoria, BC, V8P 1A1, Canada}
\altaffiltext{4}{Zentrum f\"{u}r Astronomie der Universit\"{a}t Heidelberg, Landessternwarte, K\"{o}nigstuhl 12, D-69117 Heidelberg, Germany}
\altaffiltext{5}{Kavli Institute for Astronomy and Astrophysics, Peking University, Yi He Yuan Lu 5, HaiDian Qu, Beijing, P.R. China; gherczeg1@gmail.com}
\altaffiltext{6}{Max-Planck-Institut f\"ur extraterrestriche Physik, Giessenbachstrasse 1, 85748, Garching, Germany;  simonbruderer@gmail.com}

\begin{abstract}
Stars may be assembled in large growth spurts, however the evidence for this hypothesis is circumstantial.  Directly studying the accretion at the earliest phases of stellar growth is challenging because young stars are deeply embedded in optically thick envelopes, which have spectral energy distributions that peak in the far-IR, where observations are difficult.  In this paper, we consider the feasibility of detecting accretion outbursts from these younger stars by investigating the timescales for how the protostellar envelope responds to changes in the emission properties of the central source.   The envelope heats up in response to an outburst, brightening at all wavelengths and with the emission peak moving to shorter wavelengths.  The timescale for this change depends on the time for dust grains to heat and re-emit photons and the time required for the energy to escape the inner, optically-thick portion of the envelope. We find that the dust response time is much shorter than the photon propagation time and thus the timescale over which the emission varies is set by time delays imposed by geometry. These times are hours to days near the peak of the spectral energy distribution and weeks to months in the sub-mm.  The ideal location to quickly detect  continuum variability is therefore in the mid- to far-IR, near the peak of the spectral energy distribution, where the change in emission amplitude is largest.  Searching for variability in sub-mm continuum emission is also feasible, though with a longer time separation and a weaker relationship between the amount of detected emission amplitude and change in central source luminosity.  Such observations would constrain accretion histories of protostars and would help to trace the disk/envelope instabilities that lead to stellar growth. 
\end{abstract}


\section{Introduction}
\label{sec:intro}
A star accretes most of its mass during a $\sim 0.5$ Myr period while it is deeply embedded within a dense envelope \citep{Evans2009}. In the seminal \citet{Shu1977} model of spherical protostellar collapse, the young star grows steadily from the infalling envelope, with the fixed mass-infall rate determined solely by the envelope sound speed. \citet{Kenyon1990} found, however,  that the luminosities of most protostars fall far below those expected from energy release by steady accretion over protostellar lifetimes.  They suggested that rare accretion events (outbursts) could reconcile this discrepancy.  Using large photometric samples of protostars, \citet{Dunham2010} confirmed the severity of the "luminosity problem", and found that the observed luminosities are well explained when invoking accretion outbursts, which may account for $\sim 90$\% of the total growth in stellar mass.  The protostellar luminosity function may instead be explained by competitive accretion, which increases the duration of weak accretion \citep{Offner2011,Myers2011,Kryukova2012}.

The episodic accretion hypothesis has significant support from later stages of pre-main sequence stellar evolution.  Spectacular outbursts of FUors and EXors are usually interpreted as episodes with accretion rates enhanced by factors of 100--1000 \citep[e.g.,][]{Herbig1977, Herbig2008,Hartmann1996,Aspin2010}.  Models of disks suggest that such outbursts may be triggered by the combination of the gravitational instability and magneto-rotational instability \citep{Armitage2001,Zhu2010},
although orbital evolution of binary star/disk systems \citep{Reipurth2004}, gravitational clumping \citep{Vorobyov2005}, and thermal instabilities \citep{Bell1995} could also explain these violent outbursts.  Disk instabilities are likely not limited to only these large outbursts of accretion.  For classical T Tauri stars, accretion is also unsteady on all timescales, from hours to years \citep[e.g.][]{Herbst1994,Grankin2008,Alencar2012,Costigan2012}.  Such variability is also expected for younger, more embedded sources.

Despite many well-known cases of large accretion outbursts seen at later evolutionary stages, they are difficult to directly study in the youngest protostars.  Searches for outbursts are easiest to perform in the optical and near-IR \citep[e.g.][]{McLaughlin1946,Covey2011,Morales2011}.  Although some FUors and EXors retain remnant envelopes, most of the stellar growth is thought to occur when the central protostar is too deeply embedded to detect in the optical or near-IR. Evidence for outbursts in the youngest protostars is therefore mostly indirect, with the strongest support from bolometric luminosity-temperature diagrams \citep[e.g.,][]{Dunham2010}.  Indeed, measured accretion rates from embedded objects are not significantly larger than those from CTTSs
and are $\sim2$ orders of magnitude too low to explain the growth of a
star in $10^5 - 10^6$ yr \citep{Caratti2012,Salyk2013}.
Additional support for episodic accretion can be found in periodic shocks that occur along the axis of protostellar jets, which may provide a record of the past accretion history of stars \citep[e.g.][]{Reipurth1989,Raga2002}, and from chemical abundances of Stage 0 objects \citep{Kim2011,Visser2012}.


Models of periodic outbursts are able to explain the distribution of bolometric temperatures and luminosities for protostars \citep{Dunham2010,Offner2011,Myers2011}, although models of time-dependent star formation with competitive accretion or a two-component accretion law may also explain the protostellar luminosity function \citep{Offner2011,Myers2012,Kryukova2012}.  The hypothesis of periodic outbursts can be tested directly by finding and characterizing these explosive events during the initial stages of stellar growth.  Moreover, if the accretion process onto the youngest protostars is similar to that onto older ones then similar disk instabilities or changes in the star-disk interaction should also lead to frequent changes in their accretion luminosities. Measuring the frequency and scale of large accretion events will help to assess the physics of large-scale instabilities in the disk \citep[e.g.][]{Zhu2010}.  These measurements are also necessary to determine whether accretion bursts are large enough to reduce the timescale for stellar contraction, which reduces the pre-main sequence lifetime and may introduce scatter into pre-main sequence stellar ages \citep{Baraffe2009,Hosokawa2011,Baraffe2012}.

However, the dense envelope that surrounds the youngest protostar clouds whether any variability in the central source would even be detectable.  The physical and chemical envelope models described above assume that after an accretion outburst, the envelope reaches a new equilibrium within their model timesteps, often $\sim 100$yr.  The accretion luminosity from the central protostar is reprocessed by dust in the envelope into far-IR and sub-mm photons.  Our ability to directly detect any change in the accretion luminosity from a growing protostar depends on the average travel time for energy to get absorbed by dust grains in the envelope and then re-emitted by dust at longer wavelengths, eventually escaping.  In this paper, we discuss the prospects for detecting such changes by
analyzing time-dependence in the radiative transfer of photons in the envelope of a deeply embedded protostar.

\section{Observing Envelope Variability}
\label{sec:obs}

In order to investigate the manner in which time variability of accretion onto a deeply embedded protostar affects the brightness and spectral energy distribution (SED), we begin with a simple envelope model and consider both the steady-state emission profile and the timescales required for achieving steady-state.

\subsection{Fiducial Envelope Model}
\label{sec:obs:model}

The \citet{Shu1977} inside-out collapse theory for the evolution of a protostar surrounded by an isothermal spherical envelope provides a simple semi-analytic model.  The initial density distribution $\rho_s$ in the static envelope is power-law in form, with
\begin{equation}
\rho_s = { c_s^2 \over 2\,\pi\,G} r^{-2},
\label{eqn:rhos}
\end{equation}
where $c_s(T)$ is the isothermal sound speed in the envelope, $T$ is the temperature of the gas in the envelope,  $G$ is the gravitational constant, and $r$ is the radial distance. At time $t_x$ from the start of the inside-out collapse, material inside of $R_x = c_s\,t_x$ will be infalling with a shallower density power-law such that over most of the infall it has the form
\begin{equation}
\rho_i = {3 \over 4}\, { c_s^2 \over 2\,\pi\,G} R_x^{-1/2}\,r^{-3/2}.
\label{eqn:rhoi}
\end{equation}
At the same time, the central protostellar source will have accumulated a mass
\begin{equation}
M_{\rm ps} = {c_s^2 \over G} R_x,
\label{eqn:mps}
\end{equation}
given the constant infall rate
\begin{equation}
\dot{M}_{\rm ps} = {c_s^3 \over G}.
\label{eqn:mdot}
\end{equation}
We further consider the  initial envelope mass to be $M_{\rm env}$, which we attain by truncating the outer portion of the singular isothermal sphere. Thus the outer radius of the static envelope is
\begin{equation}
R_{\rm env} = { G \over 2\,c_s^2} M_{\rm env}.
\label{eqn:renv}
\end{equation}
For our fiducial model we need to define three parameters, ($M_{\rm env}, t_x, T$), or alternatively ($M_{\rm env}, M_{\rm ps}, T$).

The luminosity $L_{\rm ps}$ produced by the protostar at the centre of the envelope can be estimated from its mass $M_{\rm ps}$ and an appeal to protostellar models such as \citet{Stahler1983}. For accreting low-mass protostars the internal radiation is augmented by the accretion luminosity $L_{\rm acc}$, which should be significant and is given as
\begin{equation}
L_{\rm acc} = { G\,\dot{M}_{\rm ps}\, M_{\rm ps} \over R_{\rm ps} },
\label{eqn:lacc}
\end{equation}
where the protostellar radius is also tabulated by \citet{Stahler1983}.  Although epidosic accretion may decrease the photospheric luminosity relative to the steady state \citep[e.g.][]{Baraffe2009}, this change from the steady state would only reduce the age of the star in our fiducial model. Combined, the steady-state luminosity from the central source is expected to be
\begin{equation}
L_{\rm tot} = L_{\rm acc} + L_{\rm ps}.
\end{equation}
The emergent spectrum of the source will depend upon the temperature of the photosphere and the contribution and shape from the accretion flow.  In the spherical envelope case, however, the extremely optically thick inner protostellar envelope effectively reprocesses all of this radiation. Thus, the emergent SED should be independent of the SED from the central region\footnote{We note that while the protostar will heat the inner region of the envelope to temperatures much higher than the isothermal value used to compute the initial density distribution, the heated gas is infalling and has little opportunity to react in response to this physical change. Thus, the derived isothermal density distribution should hold even though the gas is no longer isothermal.}.

The opacity within the envelope is provided by dust. We assume that the dust opacity follows that for thin icy mantles mixed within a gas density of $10^6\,$g\,cm$^{-3}$ \citep[commonly referred to as OH5 opacities][]{Ossenkopf1994} and that the dust mass is one percent of the gas mass. The optical depth through the envelope is dominated by the material residing at small distances from the protostar.  Thus understanding the morphology of the envelope on small scales is important. The envelope is expected to have an inner cavity surrounding the protostar due to the presence of an outflow (see discussion in \S 3.3).
As well, dust close to the protostar will evaporate when the temperature is high, removing the primary source of opacity in the most inner regions of the envelope. 

For the spherical envelope model, the exact location of this inner opacity cavity does not play a significant role in determining the emergent spectrum.  Most of the energy from the protostar is initially emitted as optical photons and then is reprocessed to much longer wavelengths before escaping the system. In this case the envelope can be effectively broken into two pieces: (1) an inner zone where the effective temperature in the envelope is such that the photons produced are {\it trapped}, and (2) an outer zone where the photons produced are {\it free} to escape. Following \citet{Hartmann1998}, the approximate location of the boundary between these zones, $R_{\rm ph}$, can be determined by requiring that at this radius the temperature be approximately that of a black body radiating such that $L_{\rm tot} = 4\,\pi\,R_{\rm ph}^2\,\sigma\,T^4$ while, at the same time, the optical depth {\it outward} through the envelope from this location for photons at the peak of the blackbody distribution is $\tau \simeq 1$. We refer this location as the effective photosphere of the envelope.

We take as our fiducial model a $1.5\,$M$_\odot$ envelope with an isothermal temperature of $10\,$K, resulting in a total size of $R_{\rm env} = 1.8 \times 10^4\,$AU. The embedded protostar is taken to have a mass of $0.25\,$M$_\odot$ and a luminosity $L_{\rm ps} = 1.2\,$L$_\odot$, reflecting an age of about $1.6 \times 10^5\,$yrs after the start of collapse.  Steady-state accretion should produce $L_{\rm acc} \sim 5\,$L$_\odot$ but as noted in the introduction, observations of deeply embedded protostars do not in general reveal such high luminosities, implying lower steady-state accretion rates. In order to test the importance of variability in the accretion rate, we consider three scenarios: $L_{\rm tot} = 1.2, 12.0,$ and $120.\,$L$_\odot$, referring to them as Quiescent (no or little accretion, equivalent to a rate $0 \times$ steady-state), $10\times$Burst (enhanced accretion, equivalent to a rate $2.2\times$ steady-state), and $100\times$Burst (very enhanced accretion, equivalent to a rate $12\times$ steady-state), respectively.   The values chosen are only meant to be illustrative of the underlying processes while the theoretical determinations below show the expected scaling for any burst luminosity. For ease of reference, the physical properties related to the fiducial model are tabulated in Table\ \ref{tab:fiducial}.

We use {\it DUSTY} \citep{Ivezic1997} to compute the equilibrium dust temperature profile throughout the envelope (Figure\ \ref{fig:dusty_temp}) for each of these cases as well as the emergent SEDs (Figure\ \ref{fig:dusty_SED}).  No external interstellar radiation field is included in these envelope models, so the outer envelope can get unrealistically cold.  We return to this issue in the Discussion, \S \ref{sec:disc}. Three aspects of the plots are immediately obvious. First, the higher the central luminosity, the hotter the envelope at all radii, with the temperature increase roughly consistent with $T \propto L^{1/4}$. Second, the higher the central luminosity, the brighter the SED at all wavelengths. Finally, the higher the central luminosity, the shorter the peak wavelength of emission. For an excellent diffusion approximation explanation of these consequences see \citet{Hartmann1998}. 

Figure\ \ref{fig:dusty_SED} suggests that the most logical place to look for evidence of protostellar variability is near the peak of the SED, at mid- to far-IR wavelengths, where the change in the brightness of the embedded protostar varies almost linearly with the change in the source luminosity. For significantly shorter wavelengths the reprocessing in the extremely optically thick interior masks any ability to observe the protostellar variation (although this might change dramatically for non-spherical envelopes as we discuss in \S\,\ref{sec:dis:model} below). At longer (millimetre) wavelengths the spectrum is dominated by the Rayleigh-Jeans tail of the emission from the outer envelope, where the bulk of the envelope mass resides.  Thus the increase in the emitted radiation at long wavelengths varies approximately linearly with the increase in the outer envelope temperature (see Figure\ \ref{fig:dusty_temp}).  However, the observability of variations in the envelope at long wavelengths is worse than it appears in Figure\ \ref{fig:dusty_SED} since the outer parts of real protostellar envelopes are heated by the external interstellar radiation field and therefore reach a roughly constant equilibrium temperature, independent of the internal luminosity. We return to this issue in the Discussion, \S \ref{sec:obs:burst}.

The peak of the SED in Figure\ \ref{fig:dusty_SED} varies from about $100\,\mu$m for the Quiescent case to $50\,\mu$m for the $100\times$Burst. For these wavelengths, the \citet{Ossenkopf1994} dust opacities are $\kappa \sim 100 - 400\,$cm$^2$\,g$^{-1}$, resulting in a required column density of dust for optical depth unity $\sim 0.0025 - 0.01\,$g\,cm$^{-2}$. Assuming a gas to dust mass ratio of 100, and taking the fiducial envelope conditions above, this suggests that the peak emission is produced at an effective photosphere radius of $R_{\rm ph} \sim 50\,$AU. For comparison, the diamond symbols in Figure\ \ref{fig:dusty_SED} denote the location of the emission-weighted $\tau = 1$ surface within each computed envelope.

\subsection{Timescales}
\label{sec:obs:time}

Three important timescales must be considered in order to determine the observability of protostellar variability from deeply embedded sources: the timescale of the embedded source luminosity variations, expected to couple directly to changes in the accretion; the timescale for heating the envelope; and the differential photon travel times within the envelope.  The timescale for variability in the luminosity of the underlying source is perhaps the most important feature.  Large increases in luminosity can occur quickly in EXOr and FUOr outbursts.  However, given our lack of a detailed model for such changes, we assume the simplest approach, that the central luminosity increases instantaneously to a burst level and then follow the effects of both the dust heating and photon travel times. Together, these processes work to spread out in time the observed luminosity jump, thus setting constraints on the types of luminosity variations capable of being observed. We will return to the issue of the timescale for the embedded source luminosity variations in \S\,\ref{sec:disc} where we discuss the feasibility of monitoring the brightness of deeply embedded protostars as a means to discern the accretion history.

\subsubsection{Heating Times}
\label{sec:obs:time:heat}

\begin{figure} 
\epsscale{0.8}
\includegraphics[angle=90,width=1.0\textwidth]{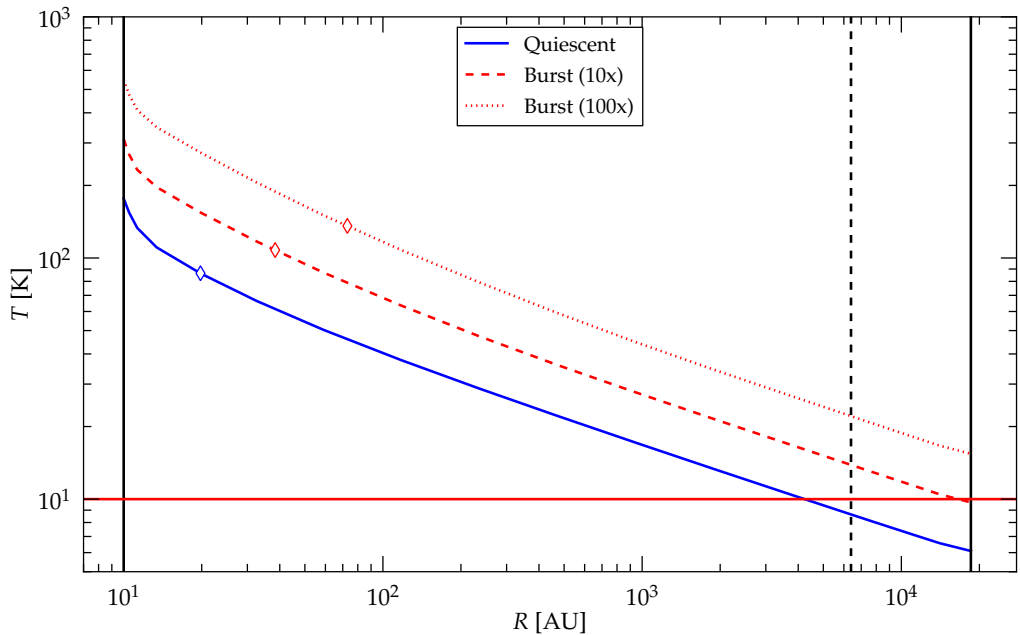}
\caption{
Dust temperature as a function of radius for our fiducial envelope (see Table\ \ref{tab:fiducial}) 
as given by the radiative transfer code {\it DUSTY} (see text). 
The quiescent case (blue solid line) assumes $L_{ps} = 1.2 L_\odot$ and no accretion luminosity.
The two different burst scenarios, 10xBurst (red dashed line) and 100xBurst (red dotted line), assume  $L_{\rm tot} =12$ and $120\, L_\odot$, respectively.
The two solid black vertical lines indicate the (externally fixed) inner evaporation radius of dust and the outer edge of the envelope, 
whereas the dashed vertical line marks the boundary between the static and infalling density regimes (see Eqn\ \ref{eqn:mps}).
The effective photosphere of the envelope is shown as a diamond marker for each model and moves outwards with increasing central luminosity.
Note that {\it DUSTY} calculates the temperature solely from internal heating, without an external radiation field.  The red horizontal line indicates 
the lower limit to the envelope temperature which we will later apply to calculate time-dependent SEDs (Figures\ \ref{fig:SED_time_10burst} and \ref{fig:SED_time_100burst}) . 
}
\label{fig:dusty_temp}
\end{figure}

\begin{figure} 
\epsscale{0.8}
\includegraphics[angle=90,width=1.0\textwidth]{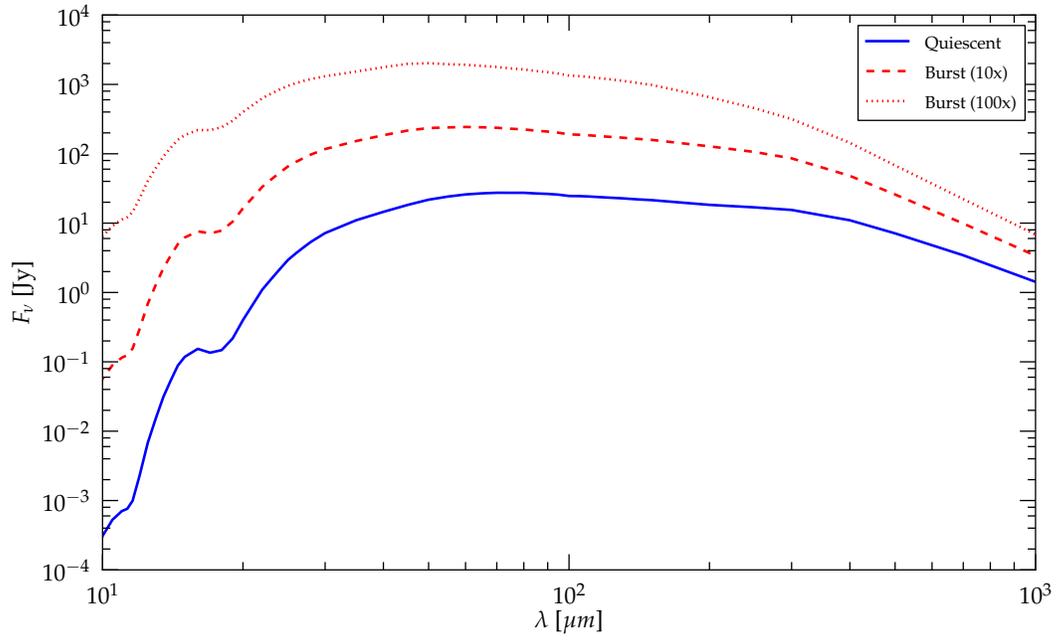}
\caption{
Equilibrium Spectral Energy Distributions as obtained from {\it DUSTY} for the envelope properties shown in Figure\ \ref{fig:dusty_temp}. 
Around the peak wavelength the brightness varies almost linearly with the total source luminosity, $L_{\rm tot}$. At long wavelengths the 
brightness varies significantly less (see text). }
\label{fig:dusty_SED}
\end{figure}

Figure\ \ref{fig:dusty_temp} shows the required change in the dust temperature as a function of position within the protostellar envelope when the luminosity of the central source increases, as expected during major accretion events. This temperature change, however, is not instantaneous as the dust retains a non-zero heat capacity and must absorb photons in order to heat up and re-emit in steady-state at the new temperature.

The timescale for heating the envelope, as a function of radius, can be effectively broken into two parts. First, in the highly optically thick part of the envelope (i.e. when $r < R_{\rm ph} $ see \S\ \ref{sec:obs:model}) a wave of energy will spread outward as fast as the dust can be heated to the new equilibrium temperature. Second, in the outer optically thin part of the envelope the heating time for the dust will be determined by a combination of the enhanced radiation field and the absorptive properties of the dust.

\subsubsubsection{Optically Thick Zone}

Ignoring for the moment the finite speed of light, the enhanced luminosity from the central source will propagate out from the centre at a rate fixed by the requirement for energy to be absorbed by the dust in order heat the dust to the new equilibrium temperature. The amount of energy deposited depends explicitly on the change in the temperature required, the amount of dust involved, and the heat capacity of the dust. Formally,
\begin{equation}
E_{\rm abs}(r) =  \int^r_{r_0} 4\,\pi\,r^2\,\rho(r)\,x\,C_d(T)\,\Delta\,T(r)\,dr,
\label{eqn:inner_heat}
\end{equation}
where $E_{\rm abs}(r)$ is the amount of energy absorbed by the dust out to radius $r$, $C_d(T)$ is the specific heat capacity of the dust per unit mass, $x=0.01$ is the fraction of envelope mass in dust, and $\Delta\,T$ is the required change in temperature. The lower bound on the integral, $r_0$ denotes the distance from the protostar where the envelope begins (or alternatively the position within the envelope where the dust ceases to be evaporated). For the present analysis we do not vary $r_0$ in response to changes in the luminosity from the central source.

For our fiducial model we have assumed that $r_0 \simeq10$\,AU under the expectation that angular momentum or magnetic fields \citep[e.g.,][]{Terebey1984, Galli1993} will produce a large degree of flattening in the inner region of the envelope. Observations of some young deeply embedded protostars suggest that circumstellar disks may form early, although it is unclear if these disks are rotationally supported \citep{Jorgensen2009,Persson2012,Tobin2012}.

The dust heat capacity depends quite strongly on the specific properties of the dust assumed \citep{Guh1989, Chase1985}, however, a simple analytic form that fits most dust to within an order of magnitude for temperatures below $\sim 500\,$K, sufficient as it will be shown for this analysis, is
\begin{equation} 
{C}_d(T) = 10\, \left({T \over 1\,{\rm K}}\right)^3\,{\rm erg\,g}^{-1}\,{\rm K}^{-1}.
\label{eqn:cd}
\end{equation}
Given the strong dependence of the dust heat capacity on temperature, the heating time within the optically thick part of the envelope will be dominated by the most inner parts (i.e. $r \gtrsim r_0$).

To gauge the importance of the heating time, we now consider the amount of energy required to bring the inner few tens of AU in our fiducial model up to the equilibrium temperature, from zero, in the quiescent phase (see Figure\ \ref{fig:dusty_temp}). The density at 10\,AU is $\rho_i(10\,{\rm AU}) \simeq 1.1 \times 10^{-15}\,$g\,cm$^{-3}$ and the envelope mass out to 10\,AU is $\simeq 10^{-5}\,$M$_\odot$ (this mass is contained in a fairly thin shell since $r \sim r_0$), the total amount of dust is one percent of this value, and the dust temperature is $\sim 100\,$K (see Figure\ \ref{fig:dusty_temp}).  Thus,
$E_{\rm abs}(10 \,{\rm AU}) \sim 2 \times 10^{35} \ $erg,
and the time to heat
\begin{equation}
t_h = E_{\rm abs}/L_{\rm ps}
\label{eqn:inner_th}
\end{equation} 
is $t_h(10 \,{\rm AU}) \lesssim 10^2 \ $s.
At 100 AU scales the envelope mass will be about 30 times larger, as we are still in the infalling portion of the envelope where $M_{\rm env} \propto r^{3/2}$, but the dust temperature required for equilibrium will be about 3 times smaller (see Figure\ \ref{fig:dusty_temp}) and the required heating per unit mass will be about 100 times smaller.  Thus, as predicted, the heating of the optically thick region will be dominated by the inner region. For our fiducial model the time to heat the entire optically thick region will be $t_h(R_{\rm ph}) \lesssim 10^2 \ $s. 

\begin{figure} 
\epsscale{0.8}
\includegraphics[angle=90,width=1.0\textwidth]{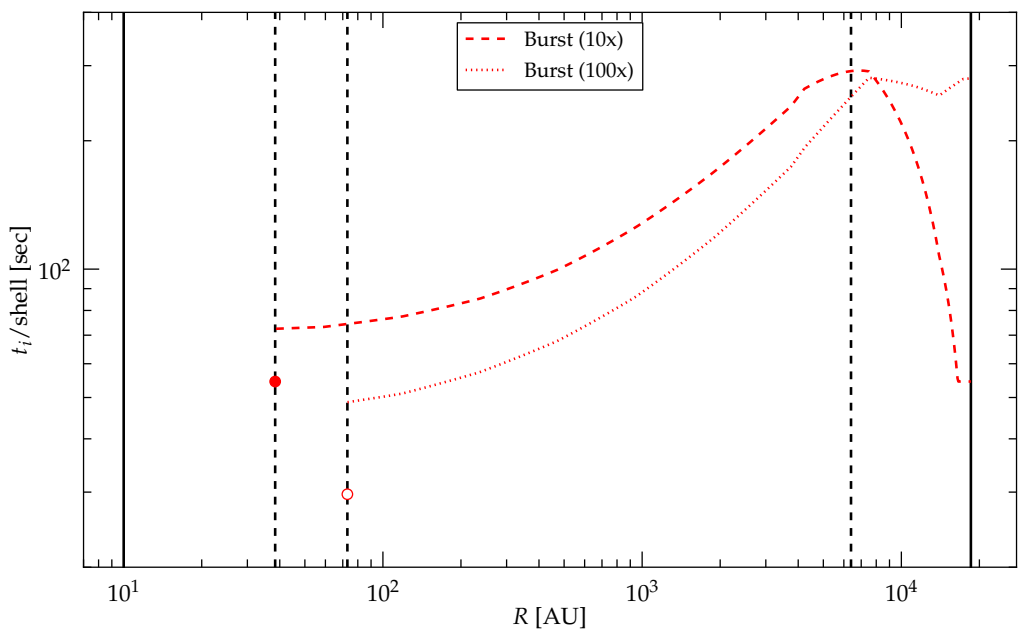}
\caption{
Heating time for the envelope as a function of radius. The red open (solid) dot indicates the integrated heating time out to
the effective photosphere, denoted by the left vertical dashed line, for the 10xBurst (100xBurst) scenario. The red dotted (dashed) line shows
 the individual heating time of the optically thin outer shells as a function of radius for the 10xBurst (100xBurst scenario).
The drop in heating time at the outermost radii is due to the imposition of the minimum dust temperature of $10\,$K  which significantly 
reduces  the temperature differential between the pre- and post-burst at the largest distances from the protostar (see Figure\ \ref{fig:dusty_temp}).
}
\label{fig:dust_heat}
\end{figure}

It is insightful to investigate the change in the envelope temperature and heating time produced by luminous bursts. The increase in the central luminosity manifests itself in an increase in the equilibrium dust temperature (see Figure\ \ref{fig:dusty_temp}). However, the increase in temperature is almost exactly $T \propto L^{1/4}$ such that the heating time $t_h = E_{\rm abs}/L_{\rm ps}$ remains effectively fixed. Figure\ \ref{fig:dust_heat} marks with circles the cumulative dust heating time for the optically thick portion of the computed envelope.

We can also estimate the effect on the heating time produced by changing $r_0$, the effective inner edge of the spherical infalling envelope. Given that $T$ varies approximately as $r^{-1/2}$ at small radii and $\rho \propto r^{-3/2}$ in the infalling zone, the integrand in Eqn.\ \ref{eqn:inner_heat} is approximately proportional to $r^{-3/2}$. Thus $E_{\rm abs} \propto r_0^{-1/2}$ and varying the inner radius of the envelope by an order of magnitude, $r_0 = 1\,$AU, only increases the heating time (Eqn.\ \ref{eqn:inner_th}) by a factor of 3. This, however, is likely to be an over-estimate of the increase in heating time with respect to $r_0$ since the heat capacity used in Eqn.\ \ref{eqn:cd} is a strong function of temperature {\it only} when the dust is below about 500\,K \citep{Guh1989, Chase1985}. By 1\,AU the equilibrium dust temperature has reached this value even in the Quiescent case. Moving the inner edge of the spherical envelope in even further will therefore produce only a very modest increase in the heating time.

Finally, it is useful to consider the time required to heat the gas as well as the dust, and the coupling time between the two components.  The heat capacity of the gas is approximately 
\begin{equation}
{C}_g \sim k/m_p,
\label{eqn:cg}
\end{equation}
where $k$ is Boltzmann's constant and $m_p$ is the mean molecular mass. Thus, ${C}_g \sim 10^8\,$erg\,g$^{-1}$\,K$^{-1}$. At 10\,AU the equilibrium dust temperature is  $\sim 100$\,K and the gas heat capacity is about 10 times the dust heat capacity, per unit mass. Taking into account that the gas mass is 100 times the dust mass, the time to heat the gas in the envelope on this scale is about 1000 times longer than that required to heat the dust, $t_g \sim 10^5\,$s, assuming there is an effective way to couple the gas to the radiation field. 

The gas to dust coupling is most likely to take place via collisions between the dust and gas. The time for a dust particle to collide with molecular hydrogen is
\begin{equation}
t_{\rm coll} = \left( n\, \sigma_d\, c_s \right)^{-1},
\end{equation}
where $n$ is the number density of molecular hydrogen, $\sigma_d = \pi\,r_d^2$ is the dust cross section, and $r_d$ is the typical dust radius. With each collision, the dust particle will lose energy to the molecular hydrogen. Given the factor of ten difference in heat capacities per unit mass, the dust particle will need to interact with an equivalent mass of molecular hydrogen equal to ten percent of its own mass in order for it to appreciably cool. Therefore, the required number of collisions is
\begin{equation}
N_{\rm coll} = 0.1 \left( {{4\,\pi \over 3}\, r_d^3\, \rho_d \over m_p} \right),
\end{equation}
where $\rho_d \simeq 2.5\,$g\,cm$^{-3}$ is the mean density of the dust. Together, the dust cooling time due to collisions with the gas is $t_c = t_{\rm coll} \, N_{\rm coll}$:
\begin{equation}
t_c  = 0.1  \left({4\,\rho_d \over 3\,\rho_i}\right)\, {r_d \over c_s},
\end{equation}
where $\rho_i = n\,m_p$. For $\rho_i(10\,{\rm AU}) \sim 10^{-15}\,$g\,cm$^{-3}$ in our fiducial model, and  assuming $r_d \sim 10^{-4}\,$cm, the gas cooling time for the dust is $t_c(10\,{\rm AU}) \sim 10^6\,s$. Given that this is much longer than the dust-photon coupling time, there is an effective separation of events, with the dust heating first and the gas reaching an equilibrium temperature much later.

Putting all of the timescales together, the optically thick region of the envelope around the deeply embedded protostar should reach an equilibrium dust temperature in $t_h(R_{\rm ph}) \lesssim 10^2\,$s, ignoring the finite speed of light which we return to in  Section \S\ref{sec:obs:time:travel}.  The gas will heat over a much longer time, $t_g(R_{\rm ph}) \sim 10^5\,$s, as it slowly draws energy from the already heated dust.

\subsubsubsection{Optically Thin Zone}

In the outer, optically thin part of the envelope the heating times are determined locally by comparing the energy input required to heat against the luminosity absorbed. Thus, for a shell at radius $r$  in the envelope, with thickness $dr$, the amount of energy that must be absorbed in order to heat the dust is
\begin{equation}
E_{\rm abs}(r) = 4\,\pi\,r^2\,\rho(r)\,x\,C_d(T)\,\Delta T(r)\, dr.
\end{equation}
Equivalently, the energy absorbed per unit time from the interior by the shell is
\begin{equation}
L_{\rm abs}(r) = L_{\rm tot}\,d\tau,
\end{equation}
where $d\tau = \rho\,x\,\kappa\,dr$ and $\kappa$ is the absorptivity of the envelope per unit dust mass, averaged over all intensity-weighted wavelengths. The heating time $t_h = E_{\rm abs}/L_{\rm abs}$ is therefore
\begin{equation}
t_h = {4\,\pi\,r^2\, C_d(T)\,\Delta T(r) \over L_{\rm tot}\,\kappa},
\label{eqn:outer_th}
\end{equation}
which is independent of both the density $\rho(r)$ and the shell thickness $dr$.

Figure\ \ref{fig:dusty_SED} shows that most of the protostellar luminosity is re-emitted near 50\,$\mu$m at the effective photosphere. The \citet{Ossenkopf1994} opacity at these wavelengths is $\kappa \sim 400\,$cm$^{2}$\,g$^{-1}$. Therefore, at a typical radial location in the envelope  $r \sim 10^3$\,AU, the dust temperature is $\sim 10\,$K, and the heating time is
$t_h(10^3\,{\rm AU}) \sim 3 \times 10^2\,$s. In Figure\ \ref{fig:dust_heat} we calculate directly this heating time from the calculated envelope properties as a function of position.

The radial dependence for a centrally heated envelope shows $t_h \propto r^2\,T^4$ which is almost linear with radius since  $T$ varies approximately as $r^{-1/4}$ at large distances.  Thus, at 10$^4$\,AU, the heating time for the fiducial model is $t_h(10^4\,{\rm AU})  \sim 3 \times 10^3\,$s (see also Figure\ \ref{fig:dust_heat}). Furthermore, during a burst the temperature in the envelope increases roughly as $T \propto L^{1/4}$.Therefore, as in the optically thick case above, the time for the dust to reach equilibrium in the outer envelope is almost independent of the internal luminosity burst strength.

In this  outer envelope environment it is also useful to consider the time required to heat the gas as well as the dust, and the coupling time between the two components. For these low temperatures the difference in the heat capacity of the dust and the gas, per unit mass, is considerable, $C_g/C_d \sim 10^4$ (see Eqns.\ \ref{eqn:cd} and \ref{eqn:cg}). At 10$^3$\, AU, the collision time between the dust and gas, however, is much longer than in the extremely dense interior as the gas density is lower by a factor of $\sim 10^3$. Compared with the interior, the time between collisions increases by $\sim 10^3$ due to the lower gas density while the number of collisions required to transfer significant heat from the dust to the gas decreases by $\sim 10^3$ due to the much lower heat capacity for dust at $10\,$K. As such the dust cooling time due to collisions with the gas remains the same; $t_c(10^3\,{\rm AU}) \sim 10^6\,s$. 

The time to heat the gas to the same temperature as the dust, however, increases dramatically. Compared with the dust heating time, the gas heating time is set by both the ratio of the heat capacity per unit mass and the ratio of total gas mass to dust mass:
\begin{equation}
t_g =   x^{-1}\,\left({C_g \over C_d}\right)\,t_h.
\end{equation} 
Substituting in the fiducial values above,
$t_g(10^3\,{\rm AU}) \sim 10^8 - 10^9\,$s.

Putting all of this together, the optically thin outer region ($r \sim 10^3\,$AU) of the envelope around the deeply embedded protostar should reach an equilibrium dust temperature on a timescale of $10^2 - 10^3\,$s, ignoring the finite speed of light (see \S \ref{sec:obs:time:travel}). The gas will heat over a much longer time, $\sim 10^8 - 10^9\,$s.  Even in steady state, thermal balance calculations indicate that the gas temperature is decoupled from and lower than the dust temperature in outer regions of cores \citep{Doty1997}.  
Nevertheless, the extreme separation of timescales for dust and gas heating seen here may have significant observational signatures (see \S \ref{sec:disc}).

\subsubsection{Photon Travel Times}
\label{sec:obs:time:travel}

In the above analysis of the heating time, we explicitly ignored the finite speed of light.   The electromagnetic propagation was assumed to be instantaneous, so that the heating time depended only on either the time to indirectly heat the dust via contact with the expanding wave of energy (optically thick deep interior) or directly heat the the dust via absorption of photons emitted from the effective photosphere (outer optically thin envelope). In this section we consider the importance of the finite speed of light in setting the timescales for observability of temperature variations in the envelope.

When a change occurs in the luminosity of the deeply embedded central source, the increased photon flux travels outward through the envelope at the speed of light. The speed with which this wave traverses the deep, optically thick interior, will be set by the slower of either the speed with which the envelope heats or the speed of light.  The dust heating time is $\sim 10^2\,$s at 10\,AU  (see \S \ref{sec:obs:time:heat}), implying a propagation speed of $v_{\rm prop} \sim 1.5 \times 10^{12}\,$cm\,s$^{-1}$, or fifty times {\it faster} than the speed of light. Thus, the actual progression of the photon flux, and the heating of the interior, should be set by the finite photon travel time. During the period when the interior optically thick part of the envelope is being heated, however,  there is no observational evidence of the heating. Only when the heating wave reaches the effective photosphere are the reprocessed photons able to escape the envelope without being reabsorbed.

\begin{figure} 
\includegraphics[angle=90,width=1.0\textwidth]{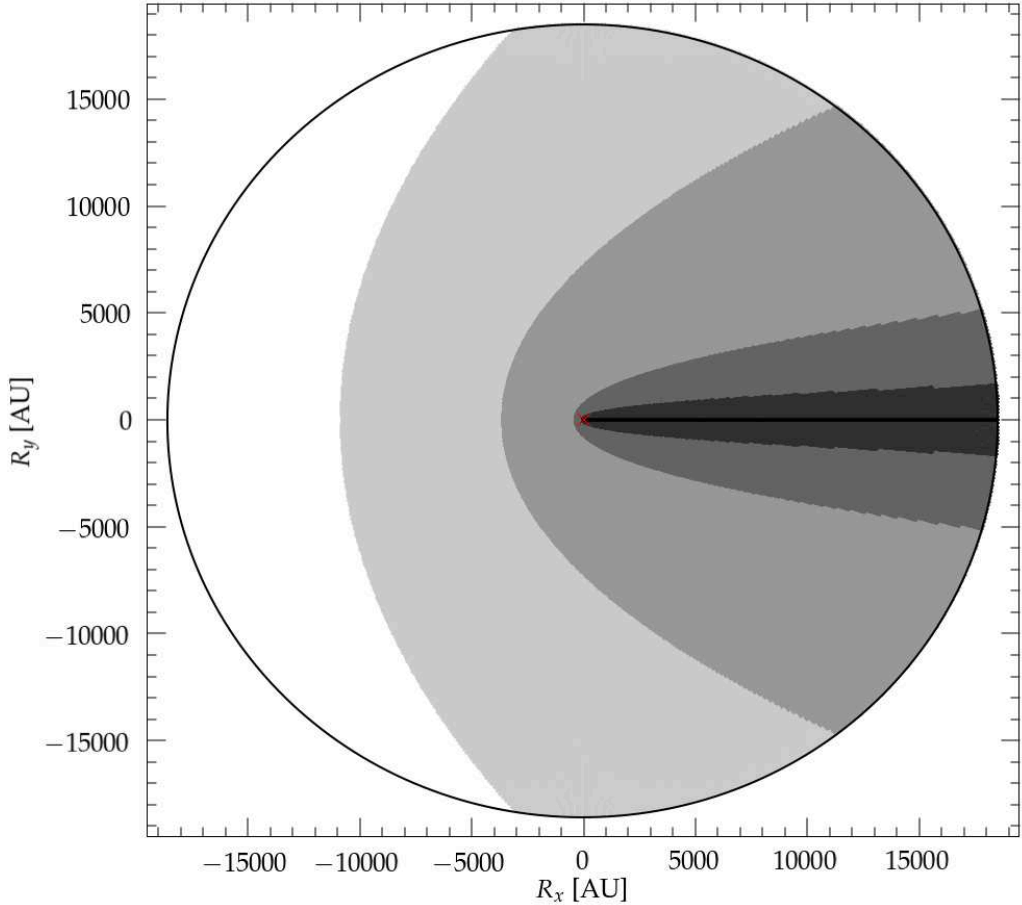}
\caption{
Schematic visualization of the observed heating process. Due to significant differential photon travel times as a function of radius and inclination angle 
in the envelope, the observed heating of the envelope is geometrically distorted (see Eqn\ \ref{eqn:photon_delay}). The grey shading indicates times of 
0h (black), 10h, 100h, 1000h, 3000h, and 6000h (white).  Time zero is defined when the first photon reaches the observer,  located in a positive x direction in the above schematic. The time delay of photons is the dominant factor in determining the time-dependent SED following a burst (see Figures\ \ref{fig:SED_time_10burst} and \ref{fig:SED_time_100burst}) . 
}
\label{fig:photon_schematic}
\end{figure}

Once the wave of heating reaches the effective photosphere, the radius at which the majority of the photons produced locally  are capable of escaping through the outer envelope, the observed SED will begin to evolve in time. For an observer at a large distance from the source, this SED variation will begin at a fixed later time set by the arrival of photons from the near-side of the effective photosphere. For the observer, however, the entire effective photosphere brightness will not change instantaneously; rather, those parts of the photosphere further from the line of sight axis will take longer to arrive due to the additional geometric path which these photons must travel and the finite speed of light (see Figure \ref{fig:photon_schematic} for a schematic diagram). For locations on this surface with an off axis impact parameter $R_b < R_{\rm ph}$, the time lag will be
\begin{equation}
\Delta t (R_b) = {R_{\rm ph} \over c}\, \left[ 1 - \left( 1 - {R_b \over R_{\rm ph}} \right) ^{1/2} \right].
\label{eqn:photon_delay}
\end{equation}

The time over which the entire effective photosphere will be observed to heat up, and increase in luminosity, will be $\Delta t \sim {R_{\rm ph} \over c}\,$, or, assuming the photospheric radius is $\sim100\,$AU, $\Delta t \sim 5\times10^4\,$s.  Thus, for an instantaneous increase in the central source luminosity the peak of the SED, which is dominated by emission near the effective photosphere, should gradually increase from pre-burst to burst conditions over a period of about half a day.

Given that the dust heating times in the outer envelope remain fast, \S\ \ref{sec:obs:time:heat}, the observed time response to the heating of the outer envelope will also be dominated by geometric effects. In general, the change in the emission of outer envelope shells should be spread in time by $\Delta t \sim {2\ R \over c}$, where the additional factor of two accounts for the fact that photons from the far hemisphere of the optically thin envelope are also able to reach the observer (see Figure \ref{fig:photon_schematic}). Since the outer shells at distances of $10^3 - 10^4\,$AU contribute significantly to longer (millimetre) wavelength emission, the timescale over which the SED evolves at these wavelengths will be weeks to months.

\section{Discussion and Complexities}
\label{sec:disc}

\subsection{Observing a Burst}
\label{sec:obs:burst}

As described in \S 2, an instantaneous increase in the central  luminosity of a deeply embedded protostar will appear  spread out in time to a distant observer primarily due to the geometric travel time of photons through the radially heating envelope (Figure \ref{fig:photon_schematic}). The first evidence of the heating will be an increase in the luminosity of the source at wavelengths near, and shorter than, the peak of the SED ($\lambda \sim 50\,\mu$m). At these wavelengths, the SED will increase from its pre-burst equilibrium to a burst equilibrium in a matter of hours. At longer wavelengths the time for the SED to reach the burst equilibrium value will be significantly delayed, weeks to months, owing to the larger geometric paths for photons in the cooler outer envelope. 

Figure\ \ref{fig:dusty_SED} reveals that the largest fractional change in the SED due to an increase in the central source luminosity occurs near the peak wavelength, where the strength of the variation is nearly identical to the variation in the burst luminosity. At longer wavelengths the difference between the pre-burst and burst SED is much less pronounced. In fact, for realistic, externally heated outer envelopes the evolution at long wavelengths is expected to be even less than that shown by the figure, since only internal heating of the envelope has been taken into account using {\it DUSTY}.

The outer portions of real protostellar envelopes are heated by the interstellar radiation field and are found to have effective temperatures between 10 and 20\,K \citep[][]{Evans2001,Rosolowsky2008}, similar to starless cores \citep[see][for a discussion of why this is the case]{Jorgensen2006}. As can be seen in Figure\ \ref{fig:dusty_temp}, where again only internal heating is included, the majority of the heating of the very outer envelope even in burst scenarios is dominated by the external radiation field. This implies that the observable changes in the SED at long wavelengths will be smaller than suggested in Figure\ \ref{fig:dusty_SED}.

Given the finite propagation of photons through the envelope, the observed SED as a function of time requires a proper time-dependent radiative transfer calculation in order to account for the variation in emission as the envelope heats. A much simpler approximation to the SED, however, is possible by assuming that the photons of interest are emitted primarily beyond the effective photosphere where the envelope is optically thin and that the timescale over which the heating occurs is short compared to the geometric travel times and can thus be assumed instantaneous (with a fixed delay).  The SED can then be estimated simply by totalling the flux emitted by each radial and azimuthal envelope zone and by assuming the attenuation is negligible. This approximation does an excellent job at reproducing the longer wavelengths where the opacity throughout the envelope is extremely low, but overestimates the strength of the emission near the peak of the SED, since this emission is emitted near the effective photosphere and should be somewhat attenuated by dust in the outer envelope. 

\begin{figure} 
\includegraphics[angle=90,width=1.0\textwidth]{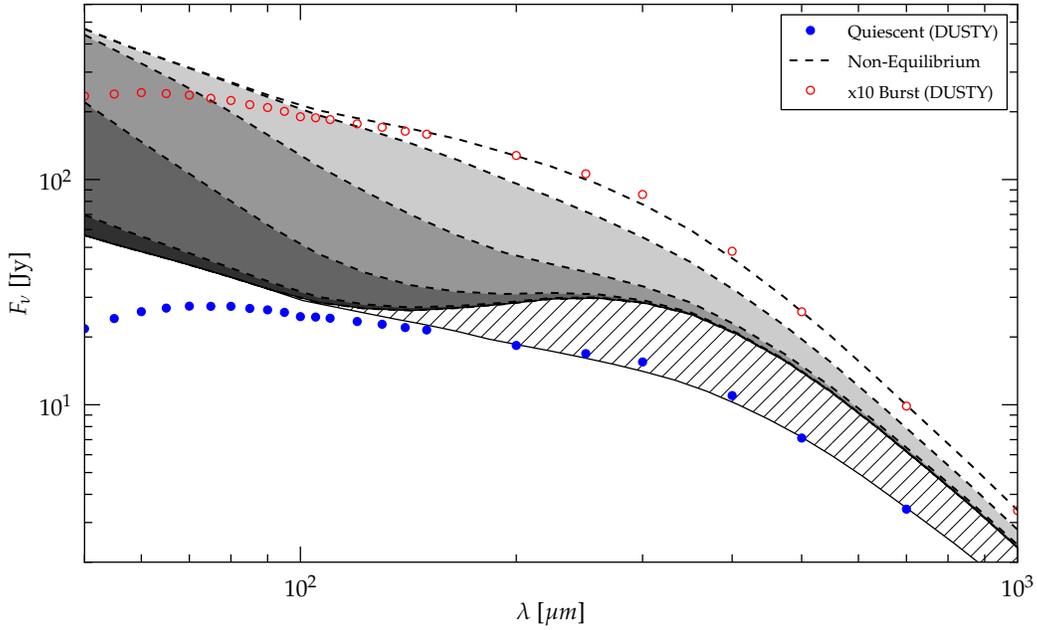}
\caption{
Time variation in the observed SED of a deeply embedded protostar after a 10xBurst increase in the source luminosity (see text). The dotted lines show the respective equilibrium SED for the quiescent case (solid blue) and the 10xBurst (empty red) as obtained with {\it DUSTY}. The grey shading follows the change in the SED over time as calculated assuming an idealized envelope, optically thin at all wavelength (see text). This assumption produces almost identical results with {\it DUSTY} for wavelengths  longer than $\sim 100\,\mu$m (with the exception of long wavelengths in the Quiescent state - see below). For significantly shorter wavelengths the optically thin assumption is incorrect and the amount of emission is overestimated. The shading corresponds to  to 0.5h (dark grey), 5h, 50h, 500h, and 5000h. After fifty hours the flux around the peak wavelength at near $100\,\mu$m has already changed by a factor of 10, whereas there has been little change at significantly longer wavelengths. 
The long wavelength region of the spectrum which experiences an extra lift, compared to the {\it DUSTY} SEDs, due to the assumption of external heating and a minimum dust temperature of $10\,$K is hatched for clarity.  
}
\label{fig:SED_time_10burst}
\end{figure}

\begin{figure} 
\includegraphics[angle=90,width=1.0\textwidth]{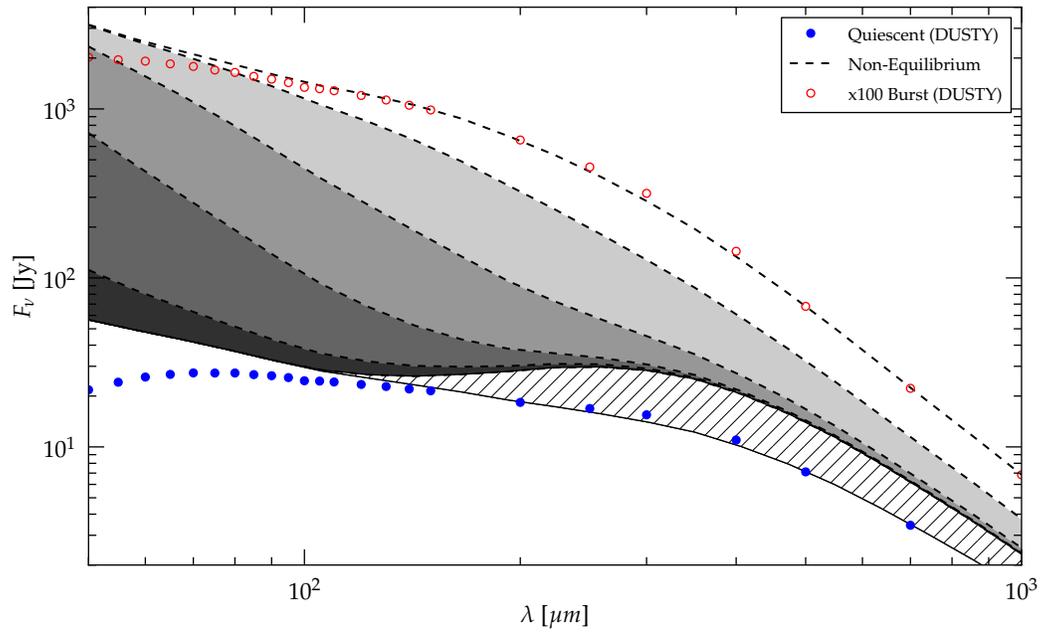}
\caption{Same as Figure\ \ref{fig:SED_time_10burst}  but for the case of a 100xBurst.
}
\label{fig:SED_time_100burst}
\end{figure}

Figures \ref{fig:SED_time_10burst} and \ref{fig:SED_time_100burst} highlight the temporal evolution of the SED for luminosity changes from Quiescent to 10$\times$Burst and 100$\times$Burst, respectively, utilizing the simplified radiative transfer  model described above.  In each plot, the initial and final equilibria SEDs are presented, assuming no external heating of the envelope and thus correspond exactly to the SEDs shown in Figure \ref{fig:dusty_SED}. Over-plotted are observed SEDs as a function of time, where the zero-point is determined by the time at which photons from the burst first escaped through the effective photosphere and reached the Earth. For the plotted time-dependent SEDs, the outer envelope temperature is fixed at 10\.K in order to more accurately reflect  real protostellar envelopes. Thus, even at early times enhanced long wavelength emission is observed, compared with the  {\it DUSTY} models, because of the change imposed on the outer envelope temperature structure.

Figures \ref{fig:SED_time_10burst} and \ref{fig:SED_time_100burst} reveal the quick increase in envelope luminosity at wavelengths near the peak of the SED\footnote{The simplification to the SED described in the previous paragraph is responsible for the strange behaviour observed at the shortest wavelengths. In particular, in the simplified radiative transfer model the shortest wavelengths reach intensities somewhat larger than expected for the burst equilibria. This occurs precisely because the attenuation of these photons through the outer envelope is not taken into account.} and the lengthened time  to reach the burst equilibria SED at longer wavelengths. The magnitude of the intensity change can be read directly from the y-axis labelling. In the case of the 10$\times$Burst, the fractional change in the intensity at mid-IR wavelengths ($\lambda \sim 50\,\mu$m)  is $\sim 10$ and occurs on a timescale of a few hours. At far-IR wavelengths ($\lambda \sim 200\,\mu$m) the fractional change in intensity is $\sim 5$ and occurs on a timescale of ten days. Finally, at sub-mm wavelengths ($\lambda \sim 500\,\mu$m)  the fractional change in intensity is $\sim 2$ and occurs on a timescale of a few months. For the 100$\times$Burst case, the fractional intensity changes increase significantly, while the timescales over which the changes are observed remain roughly constant (as anticipated in the previous section).  We note that near- and mid-IR wavelength observations might be challenging because much of the detected emission is produced in outflows rather than in the envelope \citep[e.g.][]{Jorgensen2009}, which could confuse any outburst signature.

The time-dependent SEDs shown in Figures \ref{fig:SED_time_10burst} and \ref{fig:SED_time_100burst} require that the luminosity burst of the deeply embedded protostar persist over the observed time period. If the burst is short in duration then a wave of heating followed by a wave of cooling will propagate through the envelope and the observed temperature in any radial and azimuthal envelope zone will be much more complex. Such time-dependent models are extremely interesting but beyond the scope of this paper.

\subsection{Comparing Dust and Gas Temperatures in the Envelope}
\label{sec:obs:temp}

As discussed in detail in \S \ref{sec:obs:time:heat}, the dust responds much quicker than the gas does to a burst in luminosity. For  the inner optically thick zone of the fiducial envelope model, the lag in heating the gas ($\Delta t \sim 10^5\,$s) is offset by the photon travel time ($\Delta t \sim 5\times10^4\,$s) and observing the decoupling of the gas and dust is likely to be difficult. For the outer envelope, the optically thin zone beyond the effective photosphere, the gas heating time can take many years, significantly longer than the photon travel time (months). Thus, depending on the cadence of the underlying luminosity variations, and despite the relatively high gas density $\rho_s > 10^5\,$cm$^{3}$, it is entirely possible that the outer envelope dust and gas may not attain temperature equilibrium.

\citet{Kim2011} and \citet{Visser2012} attempt to characterize the signatures of past outbursts in the chemistry of envelopes.  While some excitation of the outer disk may lag the outburst by years, many lines are produced by outflow-envelope interaction \citep{Visser2011}, where higher densities may improve the coupling between dust and gas and photochemistry would occur quickly because of the presence of outflow cavities.

Although beyond the scope of this paper, it will be extremely illuminating to connect the timescales and strengths of protostellar luminosity variations due to bursts as suggested by specific theoretical models for protostellar evolution \citep[e.g.][]{Dunham2012} with the time dependent changes in both the dust and gas temperatures in the envelope.  Such investigations could be used to define survey properties such as target cadence, continuum wavelength coverage, and which molecular lines to observe. Alternatively, a dust continuum and molecular line monitoring campaign of known deeply embedded protostars would inform the theoretical models (see \S \ref{sec:disc:surveys} below).

\subsection{Alternative Envelope Models}
\label{sec:dis:model}

The fiducial model presented above is extremely simplistic, relying on spherical symmetry and power-law radial density profiles. The results, however, are much more robust than might be initially assumed.  First, retaining the inside-out collapse model and changing only the effective temperature produces little change in the time dependent solution for the SED since the timescale is set primarily by the photon delay as a function of radius in the envelope (see Eqn.\ \ref{eqn:photon_delay}). 

For the inner, optically thick envelope this situation will change only if the mass and equilibrium temperature beneath the effective photosphere becomes extreme enough to require a significant increase in the heating time (see Eqns.\ \ref{eqn:inner_heat} and \ref{eqn:inner_th}). Given the expected range of temperatures defining low-mass protostar envelopes, $T \sim 10 - 20\,$K, this is unlikely to occur (see Eqns.\ \ref{eqn:rhos} and \ref{eqn:rhoi}). In the outer, optically thin region of the envelope, the effect of changing the envelope structure is diminished even further because the only relevant terms are the envelope temperature  and the protostellar luminosity (see Eqn.\ \ref{eqn:outer_th}), and those terms tend to counteract one another.

The situation for deeply embedded massive protostars is less clear, precisely because the physical conditions of these objects are not yet well understood. If, however, the effective temperatures are significantly higher than for low-mass analogues then the density in the inner regions will be much larger and the heating time could, in principle, become longer than the photon delay time within the optically thick zone. As well, the effective photosphere is likely to increase in size due to the enhanced envelope opacity (directly related to the density) and this will increase the time over which the peak of the SED brightens, Nevertheless, as noted above, the heating time in the outer envelope is less likely to be strongly affected by either the increase in envelope density or equilibrium temperature and the results above, perhaps scaled to larger envelope radii, should still hold.

Considering again low-mass protostars, the above commentary holds for more realistic envelope profiles, such as collapsing Bonnor-Ebert spheres \citep{Bonnor1956,Ebert1955,Myers2005}, since the enclosed mass and equilibrium dust temperatures are not significantly different than for the simpler singular isothermal sphere model. Nevertheless, while beyond the scope of this present paper, it would be useful to produce detailed time-dependent SED profiles for these often used envelope models in order to measure the finer distinctions and deviations.

Edge on disks around these deeply embedded sources will further enshroud the central protostar but will not be particularly important in determining the heating of the envelope or the observability of the changing SED unless the disk is extremely large, $R_{\rm disk} \sim R_{\rm ph}$. Alternatively, outflows driven by accretion processes carve out cavities inside the envelope \citep[e.g.][]{Snell1984,Jorgensen2007,Tobin2010}.  When the outflow axis is oriented with our line of sight, direct evidence for protostellar variability may be detected quickly.
For envelopes with outflow cavities unaligned with the observer, scattering of short wavelength IR photons off the inner wall of the outflow cavity may also provide an early beacon of burst activity. Since the only delay in observing the beginning of a burst through an optically thick envelope is due to photon propagation, the major advantage of seeing directly to the protostar is that the observed change in intensity at short IR wavelengths should be {\it almost} instantaneous, dependent only on the size of the emitting region around the protostar and expected to be $\sim R_{\rm ps}$.  Some FUOr objects still retain envelopes \citep{Zhu2008}, which must be optically-thin in our line of sight -- perhaps because of an outflow cavity.

Finally, the time-dependent SED could in principle be inverted to measure directly the density structure within an observed envelope, independent of theoretical models. In theory, combining time-sampled SEDs across a broad range of wavelengths should allow for a measure of the emission variation associated with the parabolic region traced out by the time-delay (see Figure \ref{fig:photon_schematic}).  Reverberation mapping has been successfully applied to AGN \citep[e.g.][]{Blandford1982,Peterson2004} but has not yet been applied to envelopes or disks around young stars.  Given that the expected equilibrium temperature profile in the envelope depends predominantly on the dust opacity and the embedded protostar luminosity, assuming the former \citep[e.g.][]{Ossenkopf1994} and measuring the latter should provide leverage on such an inversion. It remains uncertain as to whether the required sensitivity to these differential measurements is within the limitations of current instruments and below the non-equilibrium variations expected within the protostellar sources themselves.

\subsection{Survey Possibilities}
\label{sec:disc:surveys}

The results present in \S \ref{sec:obs} and discussed in more detail in \S \ref{sec:obs:burst}, \ref{sec:obs:temp}, and \ref{sec:dis:model} suggest that a long-term monitoring campaign of deeply embedded protostars would constrain the possible luminosity variations, and hence mass accretion rates. Indeed, while the low luminosities observed for most deeply embedded protostars suggests that mass accretion is episodic, arriving as short bursts with extreme accretion rates, very little is actually known about the cadence or the relative number of bursts with varying mass accretion amplitude. 

Thus, while it is possible that deeply embedded protostars are quiescent and non-varying for most of their evolution, with very occasional strong bursts in which the mass is deposited onto the protostar, it is equally likely that these sources are constantly varying at low mass accretion amplitudes.  With present instrumentation it should be relatively easy to monitor these sources for fractional intensity variations at or above fifty percent (and likely as low as ten percent), providing significant leverage on low amplitude mass accretion variation. Here we suggest four instruments that might provide observations which would help discriminate between the above possibilities.

Space-based imaging with the {\it Herschel Space Observatory} (Herschel), now in its last months of observations, has produced an impressive archive of star-forming molecular cloud observations using both the PACS and SPIRE instruments \citep[e.g.,][]{Andre2010,Molinari2010} -- imaging at wavelengths near the expected peak in the observed SEDs of deeply embedded protostars, where the expected variability has the largest intensity amplitude, as well as at longer wavelengths. In many cases multiple observations of these protostars have been undertaken and already variability at 70$\mu$m has been observed \citep{Billot2012}. While it is unfortunate that Herschel will not be available for a carefully defined survey in search of protostellar variability, the archive should contain a trove of useful observations for future data-mining, In particular, sources shown to significantly vary across Herschel observations should be specifically targeted for follow-up with ground-based observations.

Observing variability may be possible from the ground, but only in the driest locations.  The James Clerk Maxwell Telescope with the SCUBA-2 camera is capable of surveying large ($\sim 43$ square arcmin) regions within molecular clouds at 450\,$\mu$m and 850\,$\mu$m with excellent sensitivity \citep{Holland2006}.  As part of the telescope legacy, a large Gould Belt Survey is already underway \citep{WardT2007}, however, there are as yet no plans for systematic time-sampled observations of deeply embedded protostars.  A dedicated survey of a few small clusters of deeply embedded protostars should be both efficient and effective at uncovering variability, as each observation should take only a few minutes and differential photometry should simplify the data reduction.  Similarly, the 25m CCAT telescope under development is planned as a single-dish telescope at 5600\,m on  Cerro Chajnantor with first light instruments expected to include wide-field sub-mm cameras operating at $\lesssim 350$ $\mu$m.  Since the relevant wavelengths are longward of the peak SED of envelopes, monitoring with SCUBA-2 or CCAT should be separated in time by weeks (see Section \ref{sec:obs:burst}).  In the event that  CCAT's camera allows 200\,$\mu$m observations, the cadence of monitoring should become almost daily.

Finally, the ALMA Observatory  provides interferometric sub-mm observations and is capable of imaging the effective photosphere of nearby deeply embedded protostars. At 150 pc distance, the effective photospheres of the nearest embedded protostars will be $\sim 1\arcsec$, well within the range of spatial scales captured by ALMA. ALMA observations will therefore enable wavelength dependent, spatial dependent, and time dependent observations of protostellar variability. Additionally, ALMA can monitor both the dust continuum and individual molecular lines simultaneously, allowing for an investigation into possible deviations between the dust and gas temperature.  While using {\it ALMA} as a survey instrument to find accretion variability is unlikely, the power of ALMA could be utilized extremely effectively in following up bursts found by the JCMT or CCAT.




\section{Conclusions}
\label{sec:conc}

The growth of protostars may occur stochastically in large outbursts, while even in quiescent states the weak accretion should vary with time.  In this paper, we investigate whether and how to detect changes in accretion onto protostars.  When the central source increases in luminosity, the envelope brightens at all wavelengths and the SED  peak shifts to shorter wavelengths.  The dust quickly reaches this equilibrium in both the inner and outer envelopes. The timescales for changes in the SED are thus determined by the photon travel time and how quickly energy can propagate outward.  These changes would be easiest to detect in the mid- to far-IR, occurring on timescales of hours to days, but should also be possible at sub-mm wavelengths, occurring on timescales of days to months, with ALMA, the JCMT or CCAT.  The gas, however, reaches equilibrium much more slowly due to the slowness of the gas to dust coupling and the much larger heat capacity, per unit mass, of the gas.

\acknowledgements
\section{Acknowledgments}
\label{sec:ack}

Doug Johnstone is supported by the National Research Council of Canada and by a Natural Sciences and Engineering Research Council of Canada (NSERC) Discovery Grant.  DJ and GH would like to thank Hsien Shang and the Institute of Astronomy and Astrophysics at Academia Sinica in Taipei for hosting an enjoyable workshop in 2011, where the idea for this paper was first discussed. DJ would also like to thank the European Southern Observatory for hosting him for a month in Garching where many detailed discussions with SB took place.

\newpage

\begin{table}
\caption{Physical properties of the fiducial envelope model}
\begin{tabular}{lccr}
Property & Value & Reference\\ 
\hline
$T$ (K)& 10& initial condition\\
$t_x$ (yr)& $1.6\times10^5$ & initial condition \\
$M_{\rm env}$ ($M_\odot$)	& 1.5							& initial condition\\
$M_{\rm ps}$ ($M_\odot$)	&0.25						& Eqn\ \ref{eqn:mdot} \\
$L_{\rm ps}$ ($L_\odot$)		&1.2							& \citet{Stahler1983} \\
$L_{\rm acc}$ ($L_\odot$)	&5							& Eqn\ \ref{eqn:lacc}\\
$L_{\rm 10XBurst}$ ($L_\odot$)	&12						& - \\
$L_{\rm  100XBurst}$ ($L_\odot$)	&120					& -  \\
$R_{\rm env}$ (AU)			& $1.8\times 10^4$ 				& Eqn\ \ref{eqn:renv}	 \\
$R_x$ (AU)				& $6 \times 10^3$ 				& Eqn\ \ref{eqn:mps} \\
$R_{\rm ph}$ (AU)			& $\sim 50$					& $\tau \sim 1$\\
$\rho_i(r)$ (g\,cm$^{-3}$)		& $1.1 \times 10^{-15}  \left( r /{\rm 10\,AU} \right)^{-3/2} $& Eqn\ \ref{eqn:rhoi}\\
\hline
\end{tabular}
\label{tab:fiducial}
\end{table}

\clearpage
%
%

%
%

%
%

%
%

%
%

%
%

\end{document}